\documentclass{elsart}
\journal{New Astronomy}

\usepackage{natbib}

\usepackage{graphicx}
\usepackage{epsfig}

\usepackage{amssymb}

\begin{document}

\begin{frontmatter}



\title{Steerable wavelet analysis of CMB structures alignment}

\author{Patricio Vielva$^{1,2}$, Yves Wiaux$^3$,}
\author{Enrique Mart{\'\i}nez-Gonz\'alez$^1$ and Pierre Vandergheynst$^3$}
\address{$^1$ Instituto de F{\'\i}sica de Cantabria (CSIC - UC), Avda. Los Castros s/n, 
39005, Santander, Spain \\
$^2$ Astrophysics Group, Cavendish Laboratory, J J Thomson Avenue, Cambridge CB3 0HE, UK \\
$^3$ Signal Processing Institute, Ecole Polytechnique F\'ed\'erale de Lausanne (EPFL), 
CH-1015 Lausanne, Switzerland\\
E-mails: vielva@ifca.unican.es, yves.wiaux@epfl.ch, martinez@ifca.unican.es,
pierre.vandergheynst@epfl.ch}

\begin{abstract}

This paper reviews the application of a novel methodology for analysing the isotropy of the 
universe by probing the alignment of local structures in the CMB.
The strength of the proposed methodology relies on the steerable wavelet filtering
of the CMB signal. One the one hand, the filter steerability renders the computation of the
local orientation of the CMB features affordable in terms of computation time. 
On the other hand, the scale-space nature of the wavelet filtering allows
to explore the alignment of the local structures at different scales, probing possible different
phenomena. We present the WMAP first-year data analysis recently performed by the same authors 
(Wiaux et al.), where an
extremely significant anisotropy was found. In particular, a preferred plane was detected,
having a normal direction with a northern end position at $(\theta, \varphi) = (34^\circ, 331^\circ)$,
close to the northern end of the CMB dipole axis.
In addition, a most preferred direction was found in that plane, with a northern end direction
at $(\theta, \varphi) = (71^\circ, 91^\circ)$, very close to the north ecliptic pole.
This result synthesised for the first time previously 
reported anomalies identified in the direction of the dipole and the ecliptic poles axes. 
In a forthcoming paper (Vielva et al.), we have extended our analysis to the study of individual frequency
maps finding first indications for discarding foregrounds as the origin of the anomaly. 
We have also tested that the preferred orientations are defined by structures homogeneously distributed in the
sky, rather than from localised regions.
We have also analysed the WMAP 3-year data, finding the same anomaly pattern,
although at a slightly lower significance level.

\end{abstract}


\end{frontmatter}

\section{Introduction}
\label{sec.intro}

The release of the Wilkinson Microwave Anisotropy Probe (WMAP) first-year data in 2003 
\cite{Bennet03a} and the WMAP 3-year data in 2006 \cite{Spergel06} have implied a huge 
activity in the Cosmic Microwave Background (CMB) community during the last years.
Besides the impressive accuracy in the determination of the cosmological parameters given by 
the WMAP data analysis \cite{Spergel03, Spergel06} as well as by combining other cosmological 
data sets (e.g. \cite{Tegmark04}), a large effort was concentrated on the very challenging issue of
probing the stationary Gaussian random nature of the CMB predicted by the
standard inflationary Big-Bang model. This is a capital topic,
since the determination of the cosmological parameters that define the concordance
cosmological model relies on those assumptions.

Strong evidences (at confidence levels above $99\%$) supporting
the non-stationarity, or statistical anisotropy, of the CMB signal were established.
Analyses based on N-point correlation functions \cite{Eriksen04,Eriksen05},
local curvature \cite{Hansen04a}, local power spectra \cite{Hansen02,Hansen04b,Hansen04c,Donoghue05},
and bispectra \cite{Land05a}, suggest a north-south asymmetry maximised
in a coordinate system with the north pole at $(\theta,\varphi)=(80^{\circ},57^{\circ})$
in Galactic co-latitude $\theta$ and longitude $\varphi$, close
to the north ecliptic pole lying at $(\theta,\varphi)=(60^{\circ},96^{\circ})$.
Analyses of multipole vectors, angular momentum dispersion, as well
as azimuthal phases correlations find an anomalous alignment between
the low $l$ multipoles of the CMB, suggesting a preferred direction
around $(\theta,\varphi)=(30^{\circ},260^{\circ})$, near the ecliptic
plane and close to the axis of the dipole lying at $(\theta,\varphi)=(42^{\circ},264^{\circ})$
\cite{Copi04,Katz04,Schwarz04,Land05b,Land05c,Oliveiracosta04,Land05d,Bielewicz05}.
Galactic north-south asymmetries are also found in the analysis of
the kurtosis and the area of the wavelet coefficients of the CMB data \cite{Vielva04}.
These are mainly due to a very cold spot (the so-called \emph{cold spot}) in the southern hemisphere
(\cite{Vielva04,Cruz05,Cayon05,Cruz06a,Cruz06b} and \cite{EMG06} in this volume).
First results with the angular pair separation
method, which probes the statistical isotropy both in real and multipole
space, also seem to support those results \cite{Bernui05}. 
In opposition to the previous results, bipolar power spectra analyses are consistent with no violation
of the statistical isotropy of the universe \cite{Hajian03,Hajian05a,Hajian05b,Hajian06}.
Finally, theoretical models for an anisotropic universe are being
studied to account for the observed effects \cite{Jaffe05, Jaffe06a, Jaffe06b, Cayon06, McEwen06, Bridle06}.

The method we introduced in \cite{Wiaux06a} probes the statistical isotropy of the CMB
in a novel way: the analysis of the alignment of structures of the CMB signal. 
By searching for the directions towards which the local CMB structures are
mostly oriented, \emph{preferred directions} in the sky may be defined. The method is based
on the steerable wavelet decomposition of the CMB. On the one hand, the filter steerability renders
affordable in terms of computation time the calculation 
of the local direction of the CMB structures at each position on the sky. On the other
hand, the scale-space nature of the wavelet filtering allows the probe
of the local CMB structures alignment as a function of the scale considered. Our analysis 
synthesised for the first time previously reported anomalies related with 
the dipole and the ecliptic poles axes. In this contribution we review our first work \cite{Wiaux06a}
as well as some new results from our forthcoming paper \cite{Vielva06}, relative of the detected anomaly. 
In Section \ref{sec.method} the details of the methodology are given. 
The application to WMAP data is detailed in Section \ref{sec.apply}. 
Finally, the Conclusions are in Section \ref{sec.final}.

\section{Methodology}
\label{sec.method}

\begin{figure}
\begin{center}
\includegraphics[angle=0, width=9cm]{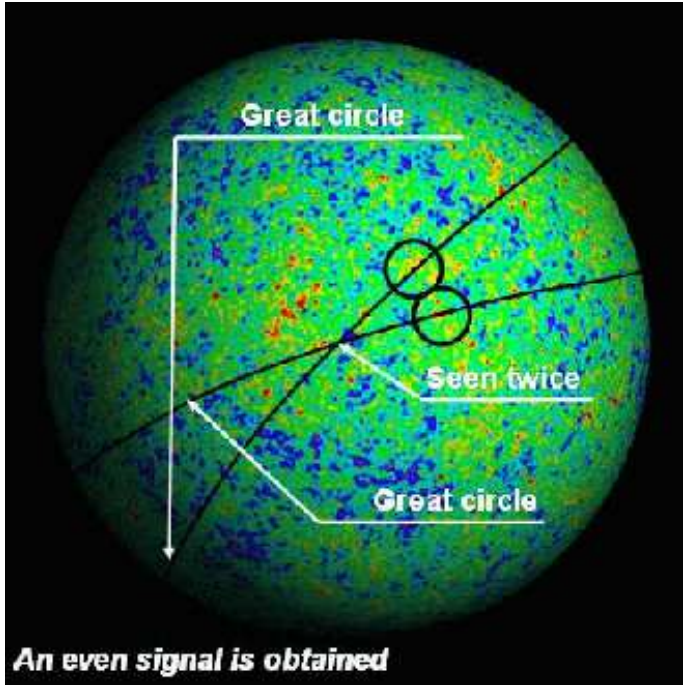}
\caption{
\label{fig.see}
Graphical illustration of the proposed methodology. The positions on the sphere towards which the local CMB
structures are aligned are defined as those lying on the great circles that pass by the structures and 
are parallel to their orientations. The signal defined in this way is even (in Cartesian coordinates) by construction.
This construction probes preferred directions in the sky towards which the CMB features are mostly oriented. 
As such, it represents a powerful analysis of the hypothesis of the universe isotropy.
This figure is a modification of one from the Max Tegmark web site (http://space.mit.edu/home/tegmark/index.html).}
\end{center}
\end{figure}

Under the cosmological principle assumption, and in the framework of the basic inflationary models, 
the CMB can be interpreted as a stationary Gaussian random field on the sphere. 
The hypothesis of stationarity, or statistical isotropy, implies that no preferred orientation
is expected. 
We propose to probe the statistical isotropy of the CMB by studying the orientation of 
its structures. The number of times a given direction in the universe is 
\emph{seen by the local CMB features} provides a unique way to determine whether there exists or not any
preferred direction on the sky towards which the CMB structures are unexpectedly aligned.

An illustration of the proposed methodology is given in Figure \ref{fig.see}. Let us focus on the two particular 
CMB structures in the middle of the two dark circles. All the positions on the sphere lying on the
great circles that pass by the selected structures and are parallel to their orientations, are said
to be \emph{seen} by the corresponding structure. As it is indicated in
the Figure, there is a particular position in the sky that is \emph{seen} by both structures. Once this procedure
is done for all the CMB structures, we end up with a signal on the sphere, allowing for a large number of
statistical analyses (e.g. angular power spectrum, geometrical statistics, Minkowski functionals, \ldots). 
This signal is, by construction, even (in Cartesian coordinates), since it is obtained by analysing the direction of the 
features, without any notion about the specific \emph{sense}.

\begin{figure}
\begin{center}
\includegraphics[angle=270, width=11cm]{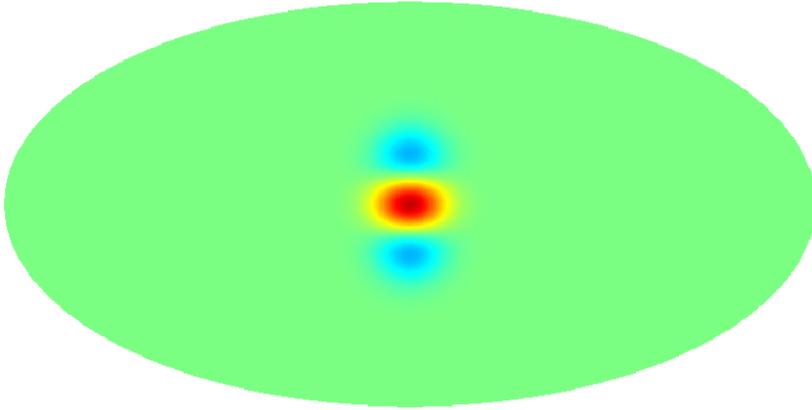}
\caption{Mollweide projection of the first member $\Psi_{\partial_{\hat{x}}^{2}, a}$ of
the \emph{second Gaussian derivative (2GD)} wavelet basis at position $\omega_0 = (\theta_0, \varphi_0) = (\pi/2, 0)$ 
and scale $a=0.19$.
\label{fig.2gd}
}
\end{center}
\end{figure}

The method is based on the steerable wavelet decomposition of the CMB signal. The reason for working
in wavelet space is twofold. First, the wavelet filtering provides the natural framework for a multi-scale analysis of the 
local CMB structures. 
This opens the possibility to probe the statistical isotropy of the CMB at different angular resolutions, 
which is crucial since a possible source of anisotropy may be due to structures of a particular size. 
Second, steerable filters render the computation 
of the orientation of the features at each position on the sky affordable in terms of the computation time
\cite{Wiaux06b, Wiaux06c}.
Steerable filters were firstly introduced on the plane by \cite{Freeman91} 
and they have been recently extended to the 
sphere by \cite{Wiaux05}. A non-axisymmetric filter is steerable if any rotation by $\chi \in[0,2\pi)$  
around itself may be expressed in terms of a finite linear combination of non-rotated basis filters. 
The reader is kindly referred to the previous references for a detailed description of the corresponding properties.
The \emph{second Gaussian derivative (2GD)} is a steerable wavelet filter, 
defined by the three following basis filters, at any scale $a>0$:

\begin{eqnarray}
\label{eq.2gd}
\Psi_{\partial_{\hat{x}}^{2}, a} \left(\theta,\varphi\right) & = & \frac{1}{a}\sqrt{\frac{4}{3\pi}}\left(1+\tan^{2}\frac{\theta}{2}\right)e^{-2\tan^{2}(\theta/2)/a^{2}}\nonumber  \left[1-\frac{4}{a^{2}}\tan^{2}\frac{\theta}{2}\cos^{2}\varphi\right]\\
\Psi_{\partial_{\hat{y}}^{2}, a} \left(\theta,\varphi\right) & = & \frac{1}{a}\sqrt{\frac{4}{3\pi}}\left(1+\tan^{2}\frac{\theta}{2}\right)e^{-2\tan^{2}(\theta/2)/a^{2}}\nonumber \left[1-\frac{4}{a^{2}}\tan^{2}\frac{\theta}{2}\sin^{2}\varphi\right]\\
\Psi_{\partial_{\hat{x}}\partial_{\hat{y}}, a} \left(\theta,\varphi\right) & = & \frac{-2}{a}\sqrt{\frac{4}{3\pi}}\left(1+\tan^{2}\frac{\theta}{2}\right)e^{-2\tan^{2}(\theta/2)/a^{2}}\nonumber \sin\left(2\varphi\right)\tan^{2}\frac{\theta}{2},
\end{eqnarray}

where the position $\omega$ at each point on the sphere is identified  by its co-latitude $\theta \in[0,\pi]$ and longitude $\varphi\in[0,2\pi)$.
In Figure \ref{fig.2gd} we plot the first member of the basis, $\Psi_{\partial_{\hat{x}}^{2}, a}$, centred
at position $\omega_0 = (\theta_0, \varphi_0) = (\pi/2, 0)$, and scale $a=0.19$.

\section{Application to WMAP data}
\label{sec.apply}

We have applied the methodology outlined in Section \ref{sec.method} to the WMAP data. The results,
obtained on the WMAP first-year data, where published in  \cite{Wiaux06a}.  In a forthcoming paper \cite{Vielva06}, 
we extend the first analysis and present the results obtained for the three-year data. In this Section
we summarise the most important results.

\subsection{Data and simulations}
\label{subsec.data}
\begin{figure}
\begin{center}
\includegraphics[angle=270, width=11cm]{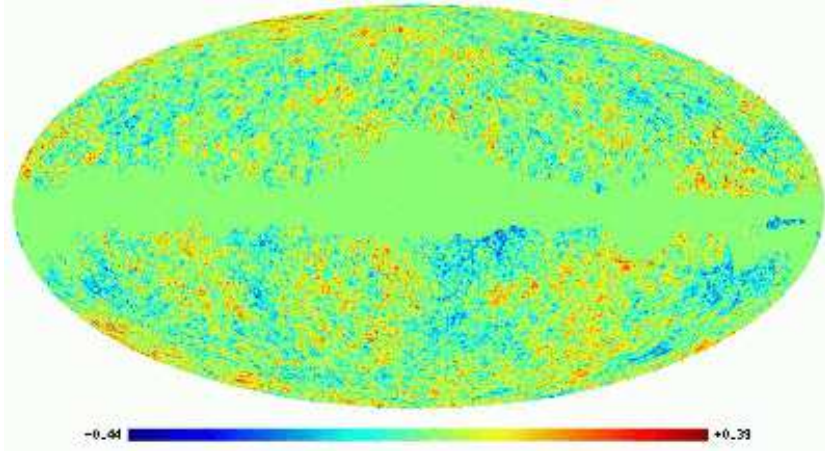}
\caption{The analysed CMB map obtained from the WMAP first-year data. This map is obtained by following the
\emph{template fits} method described in \cite{Bennet03b} and firstly used by \cite{Komatsu03} in the
WMAP Gaussianity analysis.
\label{fig.wmap_1st}
}
\end{center}
\end{figure}

Since we are interested
in probing the CMB statistical isotropy, we need to deal with a CMB map as free of foreground
emission and unknown systematics as possible. Many different CMB maps (all them obtained from the WMAP data) have 
been proposed and used during the last years. We considered the map defined  by \cite{Komatsu03}
in the WMAP Gaussianity paper. This map (the so-called \emph{template fits map}) is generated as a noise-weighted 
combination of the eight temperature maps at the Q, V and W frequency channels, previously cleaned 
from foreground emission by performing a joint multi-template fitting (see \cite{Bennet03b} for details). 
After the combination, the so-called Kp0 mask is applied to account for diffuse Galactic emission close to the
Galactic plane as well as for discarding bright point sources. Finally, the residual monopole and dipole are subtracted outside
the mask. The resultant map is presented in Figure \ref{fig.wmap_1st}. The WMAP data are
given in the HEALPix pixelization \cite{Gorski05} at resolution parameters $N_{side} = 512$ and  $N_{side} = 1024$.
However, we have performed our analysis on maps downgraded at $N_{side} = 32$, where the signal-to-noise ratio
is extremely high. At that resolution, the sphere is defined by $N_{pix} = 12288$ pixels, with an angular
resolution of 1.8$^\circ$. 

In order to confront the obtained results with the expected isotropic behaviour, we have performed 10000 simulations.
For each simulation, a stationary Gaussian CMB realisation is generated from the theoretical angular power spectrum $C_\ell$ computed
with CMBFAST \cite{Seljak96}, using the cosmological parameters given by the WMAP best-fit cosmological model.
Each receiver is simulated by convolving the simulated map with the corresponding beam window function and by
adding an anisotropic Gaussian noise realisation of the particular noise level. As for the data, the eight maps are
linearly combined using weights defined by the noise amplitude per pixel. The Kp0 mask is applied, before removing the residual monopole
and dipole.

\begin{figure}
\begin{center}
\includegraphics[angle=270, width=11cm]{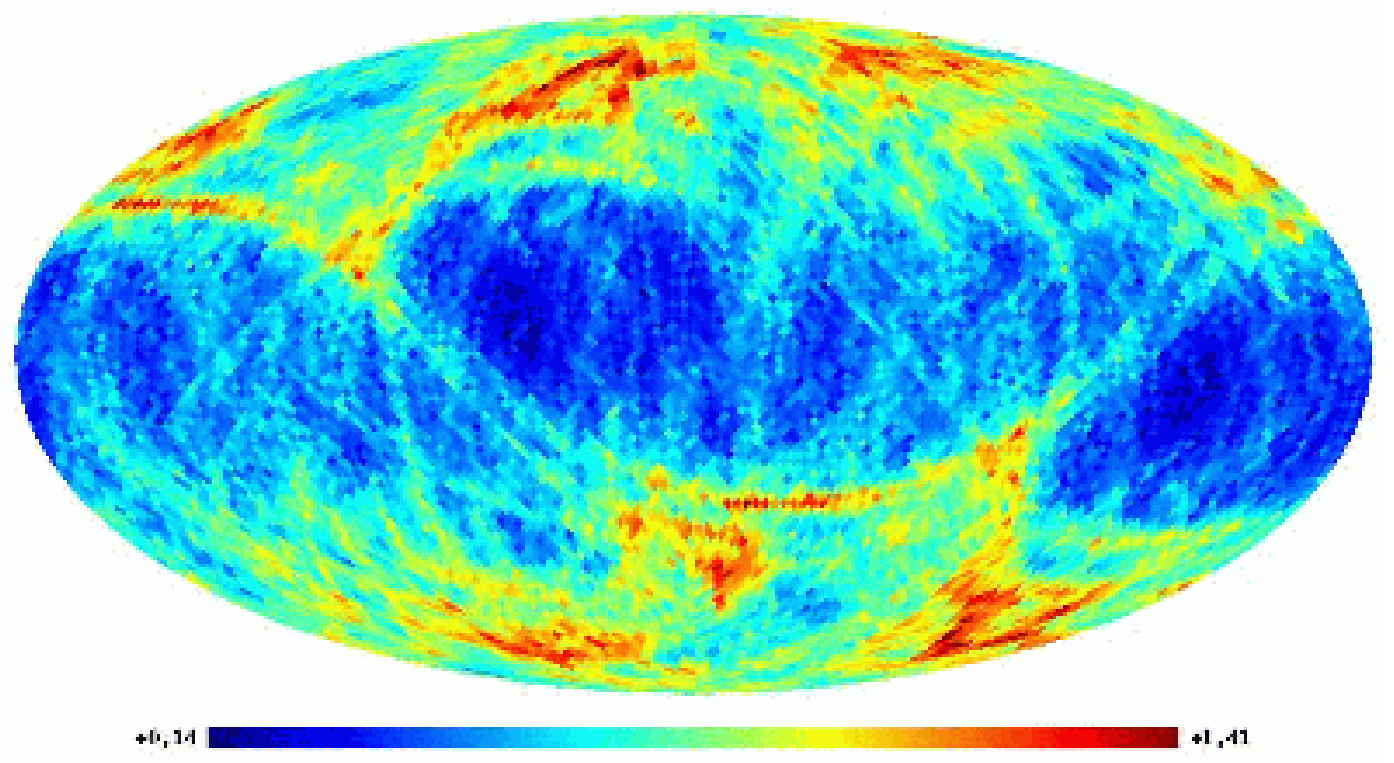}
\includegraphics[angle=270, width=11cm]{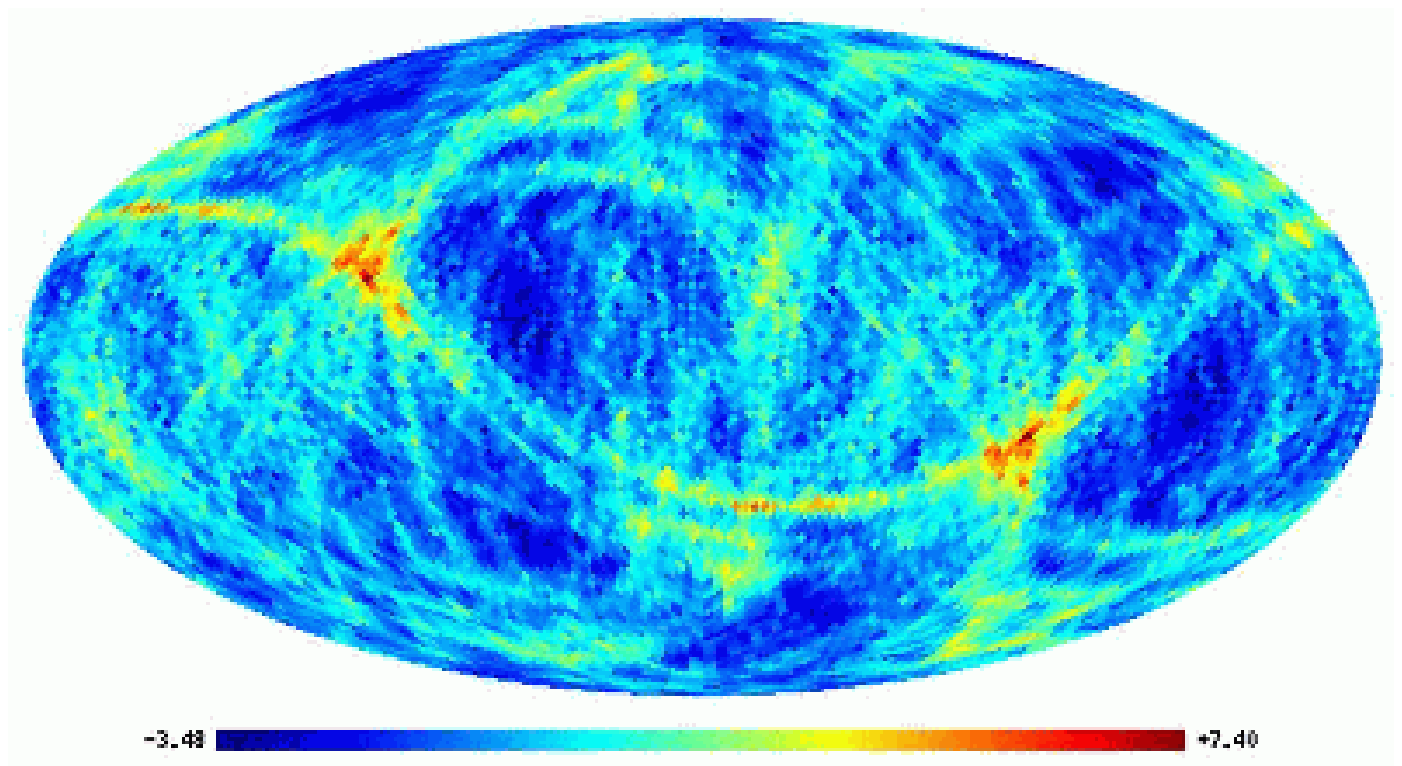}
\caption{
\label{fig.weights}
Top panel: Map of total weights $D_{a_3}\left(\omega\right)$ resulting from 
the WMAP first-year data at scale $a_3$ corresponding to an angular size of 
8.3$^\circ$. Bottom panel, the same map, but expressed in
$\sigma_{a_3}$ units (estimated from the 10000 simulations).
}
\end{center}
\end{figure}

\subsection{Results}
\label{subsec.results}

We have explored the CMB statistical isotropy at twelve different scales. For a given scale $a$ of the wavelet, the angular size
of the 2GD wavelet is defined as twice the dispersion of the corresponding Gaussian. The twelve
scales considered, correspond to angular sizes ranging from 5 to 50$^\circ$. By convolving the CMB map with
the wavelets, the pixels close to the mask are going to be very affected by the zero value of the mask.
For that reason, at each scale, an extended exclusion mask
$M_a$ is defined in order to avoid highly \emph{contaminated} pixels close to the mask.
For each pixel outside the mask in the sky and for each scale, the orientation for which the wavelet coefficient is maximum in absolute
value is selected and the corresponding absolute value is retained. This procedure selects the local wavelet
orientation that best matches the orientation of the local structure of the signal at each point.

As explained in \cite{Wiaux06a}, the total weight $D_a(\omega)$ is a weighted measure of the 
number of times a given pixel $\omega$ in the sky is \emph{seen} by local features at
a given scale $a$ in the original signal.
It is defined as:

\begin{equation}
\label{eq.tw}
D_{a}\left(\omega\right)=\frac{1}{A}\sum_{c=1}^{N_{cros}(\omega)}\vert W_{\Psi}^{F}\left(\omega_{0}^{(c)},\chi_{0}(\omega_{0}^{(c)}),a\right)\vert,
\end{equation}

where $W_{\Psi}^{F}(\omega_{0},\chi_{0}(\omega_{0}),a)$ is the maximum value of the wavelet coefficient at the
position $\omega_0$ (that is achieved at orientation $\chi_0 \in [0, 2\pi)$) and
the factor $A=LN_{pix}^{-1}\sum_{\omega_{0}\notin M_{a}}\vert W_{\Psi}^{F}(\omega_{0},\chi_{0}(\omega_{0}),a)\vert$
of normalisation defines a mean total weight in each direction equal
to unity for isotropic CMB simulations without mask: $N_{pix}^{-1}\sum_{\omega\in S}D_{a}(\omega)=1$.
The quantity $L=4N_{side}$ stands for the number of points on a great circle on a HEALPix grid.

\begin{figure}
\begin{center}
\includegraphics[angle=270, width=11cm]{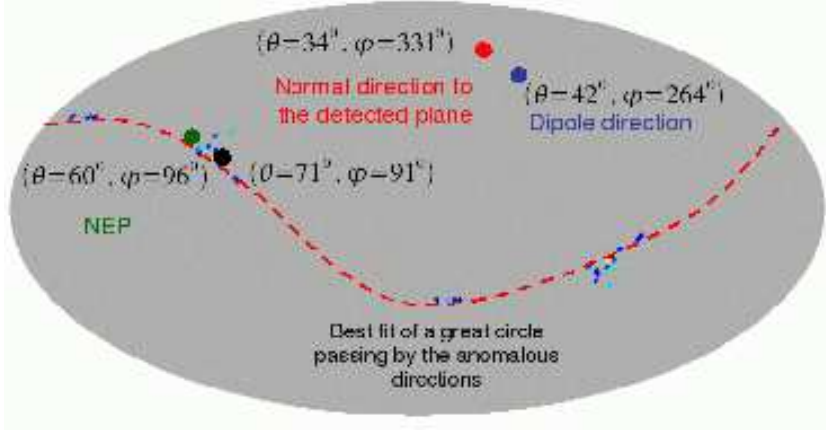}
\caption{The 20 anomalous directions (pairs of opposite points) with total 
weights higher than in any of the 10000 simulations are shown. 
They define a preferred plane with a perpendicular northern end position at  
$(\theta,\varphi)=(34^{\circ},331^{\circ})$, very close to the CMB dipole axis at 
$(\theta,\varphi)=(42^{\circ},264^{\circ})$ and to the so-called \emph{axis of evil}.
A preferred direction in the preferred plane is identified with a northern end position at  
$(\theta,\varphi)=(71^{\circ},91^{\circ})$,
which is extremely close to the north ecliptic pole
at $(\theta,\varphi)=(60^{\circ},96^{\circ})$.
\label{fig.anomalies}
}
\end{center}
\end{figure}

In the top panel of Figure \ref{fig.weights} we present the signal $D_{a_3}\left(\omega\right)$ obtained from the
analysis of the WMAP first-year data. The angular size corresponding to the scale $a_3$ is 8.3$^\circ$. It is clear that the zone 
around the Galactic plane is less \emph{seen} than the rest of the sky. It is an obvious result, 
since a pixel has more probability to be \emph{seen} from nearby pixels than from distant ones. 
Hence the exclusion of pixels by the mask implies that, even in an isotropic universe, 
the pixels close to the mask are less \emph{seen} than the others.
Obviously, what is much more interesting is to
represent the previous signal as normalised, at each point, by the
standard deviation (sigma) estimated from the 10000 simulations, i.e.
in $\sigma_a$ units. This normalisation notably cancel out all the \emph{spurious anisotropies}, like the one due to the mask. 
This map is
presented in the bottom panel. As it can be seen, there is a very particular pattern with many great circles crossing
the equatorial poles. Obviously, the distribution of  $D_{a_3}\left(\omega\right)$ is not Gaussian, and, therefore,
the number of sigmas is not a direct measurement of the probability of each direction, although it is quite
useful for a visual illustration. In any case, the probability associated 
with  each direction in the sky
(pairs of opposite points) can be easily computed from the simulations. In particular, there are 27 directions
that present values of $D_{a_3}\left(\omega\right)$ anomalous at 99.99\% (i.e., at maximum, only one simulation has
a larger value). Even more, just one simulation has at least 27 directions above 99.99\% at scale $a_3$.

Among the previous 27 directions, 20 present values of  $D_{a_3}\left(\omega\right)$ 
larger than in any of the 10000 simulations. 
These directions (Figure \ref{fig.anomalies}) are also identified as being anomalous at more than $99.99\%$
(a 100\% statement being prohibited by the finite size of our sample of simulations).
These anomalous directions are shown in Figure \ref{fig.anomalies}.
They are mostly concentrated in two clusters of 4 and 16 directions. The mean direction\footnote{The mean direction is
obtained by weighting each direction by its corresponding total weight.} of the latter 
has a northern end at  $(\theta,\varphi)=(71^{\circ},91^{\circ})$, which is extremely close to the north ecliptic pole
at $(\theta,\varphi)=(60^{\circ},96^{\circ})$. But even more, the two clusters seem to lie on a great circle, hence defining 
a preferred plane in the sky. The direction perpendicular to this preferred plane has a
northern end position at $(\theta,\varphi)=(34^{\circ},331^{\circ})$
which is very close to the CMB dipole axis at $(\theta,\varphi)=(42^{\circ},264^{\circ})$ and to the so-called \emph{axis of evil}.
Let us emphasise that also at scale $a_4$ (around 10$^\circ$ of angular size) the CMB presents the same anomaly pattern. In this
case there are 11 anomalous directions with a total weight  $D_{a_4}\left(\omega\right)$ larger than the one
obtained with any of the 10000 simulations.

\begin{figure}
\begin{center}
\includegraphics[angle=270, width=11cm]{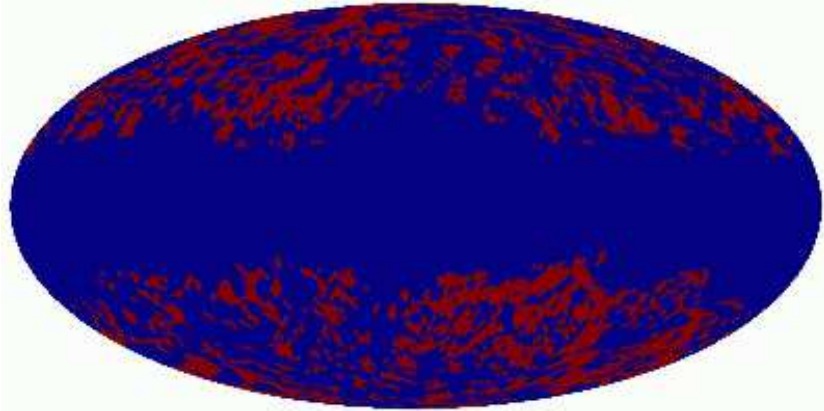}
\caption{
\label{fig.origin}
The pixels that are aligned towards any of the 20 directions anomalous at more than $99.99\%$ in the WMAP first-year data are shown (in dark red).
They are homogeneously distributed in the sky, which seems to indicate that the source of the detected anomaly is not
localised in a specific region of the sky.
}
\end{center}
\end{figure}

\begin{figure}
\begin{center}
\includegraphics[angle=270, width=11cm]{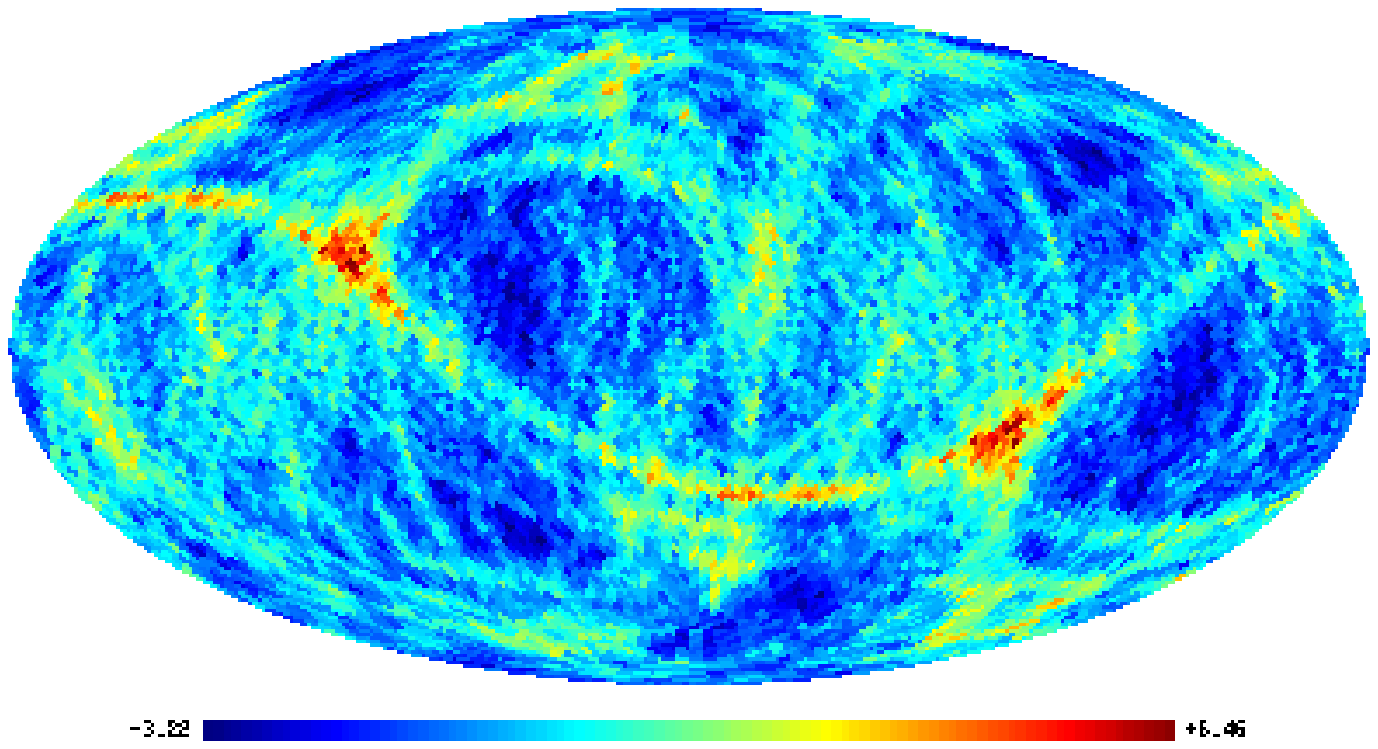}
\includegraphics[angle=270, width=11cm]{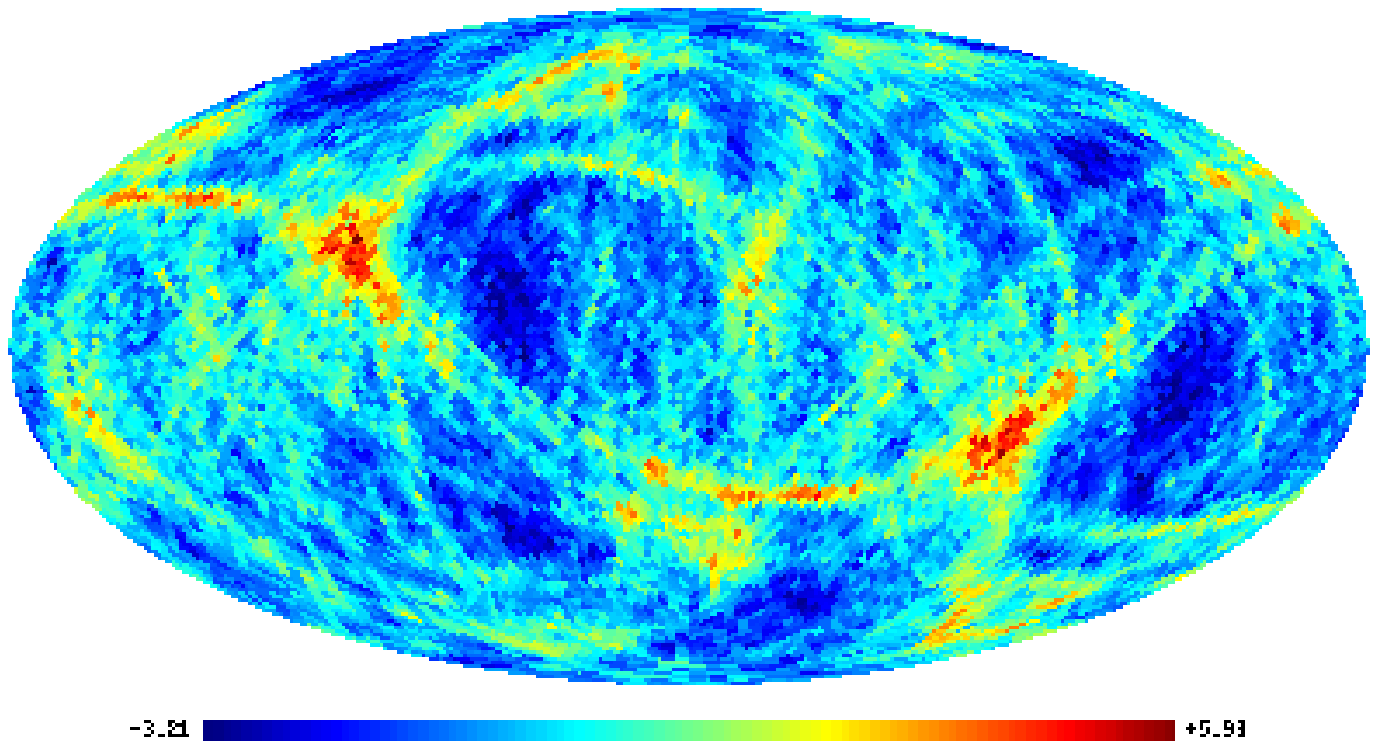}
\includegraphics[angle=270, width=11cm]{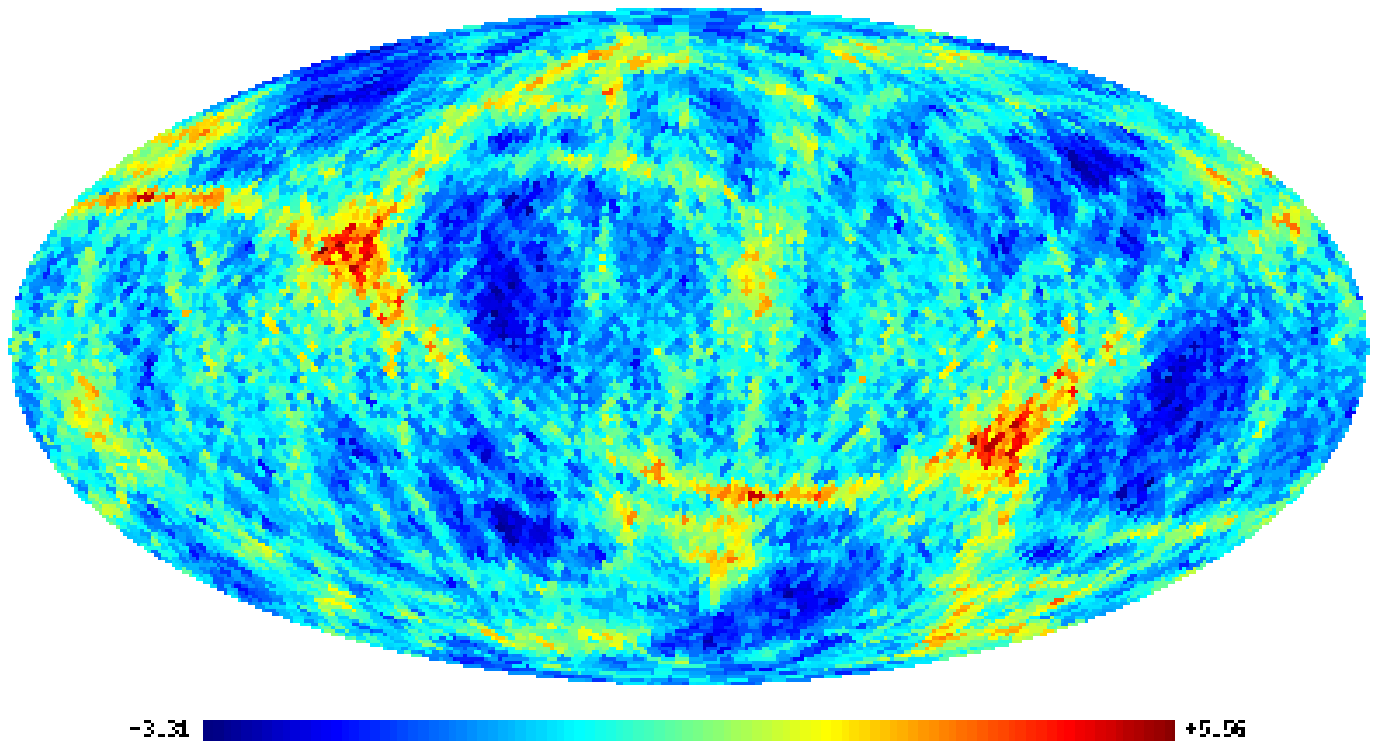}
\caption{
\label{fig.channels}
The  total weight $D_{a_3}\left(\omega\right)$ are plotted in $\sigma_{a_3}$ units for the Q (top), V (middle) 
and W (bottom) WMAP frequency maps. The obtained pattern is the same for all of them and is the same as the one already shown in
Figure \ref{fig.anomalies}. This fact seems to discard the foregrounds as responsible for the anomaly.
}
\end{center}
\end{figure}

In a forthcoming paper \cite{Vielva06}, we investigate possible origins for this anomaly.
For instance, one interesting question is to investigate whether this anomalous detection is due to some structures spatially localised or,
on the contrary, the origin for such anomaly is homogeneously distributed in the sky. To answer that question, we have plotted in
Figure \ref{fig.origin} the pixels that are aligned towards any of the 20 anomalous direction. One can infer that the
latter hypothesis is the most plausible one.

\begin{figure}
\begin{center}
\includegraphics[angle=270, width=11cm]{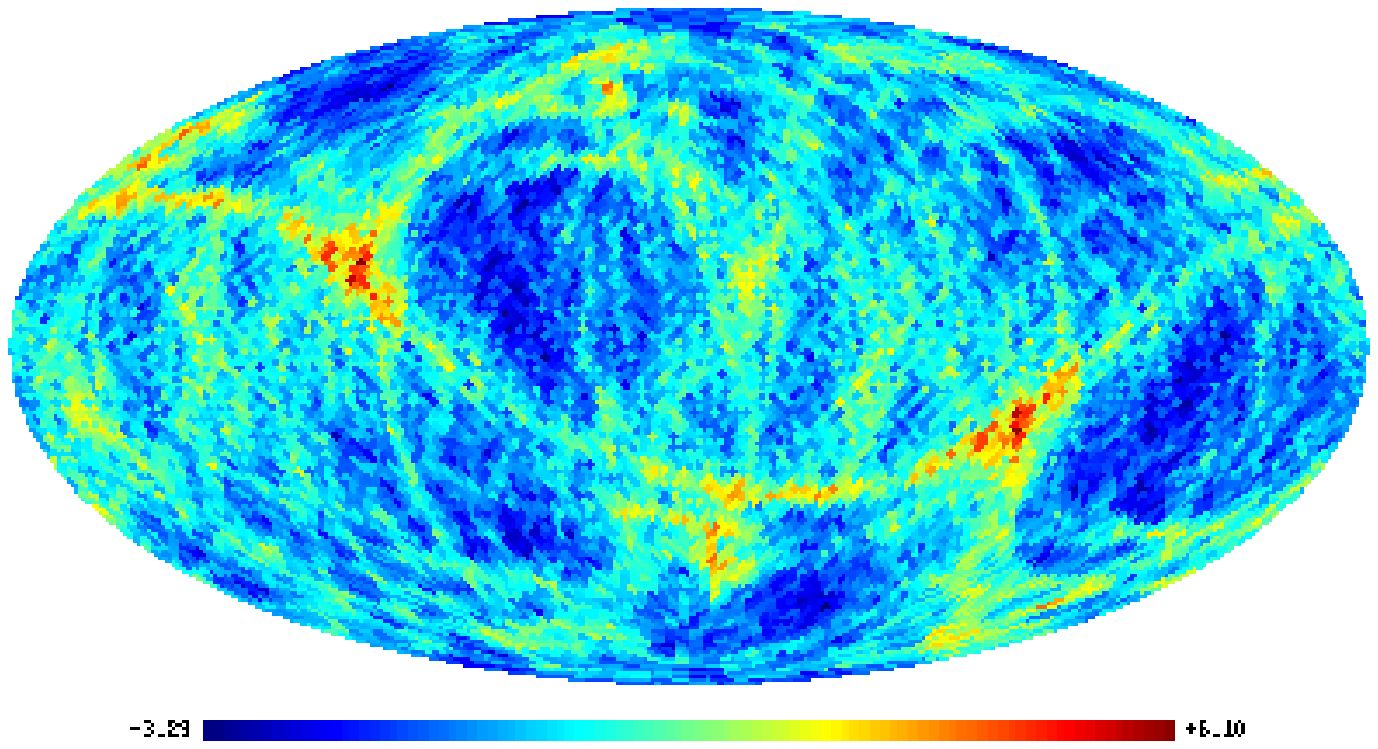}
\caption{
\label{fig.anos}
Map of total weights $D_{a_3}\left(\omega\right)$ resulting from 
the WMAP 3-year data at scale $a_3$ in $\sigma_{a_3}$ units.
The same pattern as in the first-year data analysis is observed, although the significance
of the anomaly has slightly decreased.
}
\end{center}
\end{figure}

We are also analysing whether the origin of the anomaly could be caused by a residual foreground contribution on the CMB map.
If so, one should expect to find a frequency dependence for the anomaly, as the foregrounds themselves are frequency dependent. 
In Figure \ref{fig.channels} we plot the $D_{a_3}\left(\omega\right)$ signal, in $\sigma_{a_3}$ units,
obtained for each of the Q, V and W WMAP 
frequency channels. The pattern already shown in Figure \ref{fig.anomalies} is as well observed on each one of the analysed frequencies,
indicating that the anomaly seems to be not caused by foregrounds.

Finally, the preliminary analysis of the WMAP 3-year data
also seems to indicate the same structure of anomalies at the same scales. For comparison
with the results for the first-year data (Figure \ref{fig.weights}, bottom panel), in
Figure \ref{fig.anos} we present the $D_{a_3}\left(\omega\right)$ in $\sigma_{a_3}$ units
for the 3-year data.
The same pattern can be recognised in both data sets, although for 
the 3-year data the significance of the anomaly
seems to be slightly smaller than in the first-year data: whereas in the first release
27 positions were anomalous at 99.99\% (i.e., at maximum, only one simulation has
a larger value), in the second one, only 15 positions had a 
total weight with the same significant. The number of simulations having at least 15
positions at that significance or above is 42.

\section{Conclusions}
\label{sec.final}

This contribution reviews a novel methodology for probing the universe isotropy through the
statistical isotropy of the CMB. It 
relies on the analysis of the alignment of local features in the CMB proposed 
by the authors in  \cite{Wiaux06a}. The power of the method
is based on the use of steerable wavelets. 
On the one hand, steerability makes feasible the computation of the
local orientation of the CMB features which is otherwise unaffordable in terms of computation times.
On the other hand, the steerability renders accessible the computation of the local
orientation of the CMB features, which is otherwise unaffordable in terms of
computation times. On the other hand, the scale-space 
nature of the wavelet scheme allows
to explore the alignment of the local structures at different scales, probing possible different
phenomena. 

In the first application of the method \cite{Wiaux06a} to the WMAP first-year data, we found an
extremely significant anisotropy. 20 anomalous directions were
identified in the sky at a very high confidence level ($\geq 99.99\%$) at a scale $a_3$ corresponding
to an angular size of 8.3$^\circ$, corresponding to a multipole range between $\ell = 11$
and $\ell = 27$. The 20 anomalous directions where located in two clusters
of 4 and 16 directions respectively. The mean position of the biggest one is at  
$(\theta, \varphi) = (71^\circ, 91^\circ)$, very close to the north ecliptic pole
$(\theta, \varphi) = (60^\circ, 96^\circ)$. In addition, the two clusters are aligned in the
same plane, with a northern end normal direction pointing towards 
$(\theta, \varphi) = (34^\circ, 331^\circ)$,
very close to the  CMB dipole axis $(\theta, \varphi) = (42^\circ, 264^\circ)$ 
and the so-called \emph{axis of evil}.
This result synthesised for the first time previously
reported anomalies identified towards the dipole and the ecliptic poles axes. 

We are currently analysing the origin of such an anomaly \cite{Vielva06} as well as its significance
in the WMAP 3-year data. The second WMAP data release provides a pattern of total weights compatible with the
one obtained with the first one. However, the significance of the detection seems
to be slightly lower.
We are investigating whether this anomalous detection is due to some structures spatially localised or,
on the contrary, the origin for such anomaly is homogeneously distributed in the sky. 
Our analysis indicates that the latter is the most plausible option.
We are also analysing if the origin of the anomaly could be caused by a residual foreground contribution on the CMB map.
If so, one should expect to find a frequency dependence for the anomaly, as the foregrounds themselves are frequency dependent.
The same pattern of total weights has
been found in the individual Q, V and W frequency channels of the first-year data. This might discard foregrounds as
a possible origin of such anomaly. 

Other possible sources for explaining the anomaly are being considered. Firstly, the coincidence of the preferred
direction (towards the ecliptic poles and the CMB dipole axes) naturally suggest possible unknown systematic effects
\cite{Gordon05, Freeman06}. Among them, errors in the beam pattern reconstruction are being studied. It is also
interesting to note that the angular size of the mesh of the WMAP scan pattern defined by the combination of the 
spin and precession of the satellite is of the order of several degrees \cite{Bennet03c}.
Second, the angular size at which the anomaly is detected is compatible with the size of primary CMB anisotropies due
to topological defects such as texture fields \cite{Turok90} or secondary anisotropies due to the Rees-Sciama effect
\cite{EMG90}. Alignment mechanisms \cite{Gordon05, Rakic06} were recently proposed which might be generalised to
such structures.

The authors acknowledge the use of the LAMBDA archive, and of the
HEALPix and CMBFAST softwares. YW was supported by the Swiss and
Belgian National Science Foundations. PV and EMG were supported
by the Spanish MEC project ESP2004-07067-C03-01.


\begin{thebibliography}{99}
\frenchspacing

\bibitem{Bennet03a}C. L. Bennet \emph{et al.}, Astrophys. J. Suppl. \textbf{148}, 1
(2003).
\bibitem{Spergel03}D. N. Spergel \emph{et al.}, Astrophys. J. Suppl. \textbf{148}, 175
(2003).
\bibitem{Spergel06}D. N. Spergel \emph{et al.}, submitted, preprint LAMBDA web site
(2006).
\bibitem{Tegmark04}M. Tegmark \emph{et al.}, Phys. Rev. D \textbf{69}, 103501
(2004).
\bibitem{Eriksen04}H. K. Eriksen, F. K. Hansen, A. J. Banday, K. M. G\'{o}rski, and
P. B. Lilje, Astrophys. J. \textbf{605}, 14 (2004).
\bibitem{Eriksen05}H. K. Eriksen, A. J. Banday, K. M. G\'{o}rski, and P. B. Lilje, Astrophys.
J. \textbf{622}, 58 (2005).
\bibitem{Hansen04a}F. K. Hansen, P. Cabella, D. Marinucci, and N. Vittorio, Astrophys.
J. Lett. \textbf{607}, L67 (2004).
\bibitem{Hansen02}F. K. Hansen, K. M. G\'{o}rski, and E. Hivon, Mon. Not. R. Astron.
Soc. \textbf{336}, 1304 (2002).
\bibitem{Hansen04b}F. K. Hansen, A. J. Banday, and K. M. G\'{o}rski, Mon. Not. R. Astron. Soc.
\textbf{354}, 641 (2004). 
\bibitem{Hansen04c}F. K. Hansen, A. Balbi, A. J. Banday, and K. M. G\'{o}rski, Mon. Not. R. Astron.
\textbf{354}, 905 (2004).
\bibitem{Donoghue05}E. P. Donoghue and J. F. Donoghue, Phys. Rev. D \textbf{71}, 043002
(2005).
\bibitem{Land05a}K. Land and J. Magueijo, Mon. Not. R. Astron. Soc. \textbf{357}, 994
(2005).
\bibitem{Copi04}C. J. Copi, D. Huterer, and G. D. Starkman, Phys. Rev. D \textbf{70},
043515 (2004).
\bibitem{Katz04}G. Katz and J. Weeks, Phys. Rev. D \textbf{70}, 063527 (2004).
\bibitem{Schwarz04}D. J. Schwarz, G. D. Starkman, D. Huterer, and C. J. Copi, Phys. Rev.
Lett. \textbf{93}, 221301 (2004).
\bibitem{Land05b}K. Land and J. Magueijo, Mon. Not. R. Astron. Soc. \textbf{362}, L16
(2005).
\bibitem{Land05c}K. Land and J. Magueijo, Mon. Not. R. Astron. Soc. \textbf{362}, 838
(2005).
\bibitem{Oliveiracosta04}A. de Oliveira-Costa, M. Tegmark, M. Zaldarriaga, and A. Hamilton,
Phys. Rev. D \textbf{69}, 063516 (2004).
\bibitem{Land05d}K. Land and J. Magueijo, Phys. Rev. Lett. \textbf{95}, 071301 (2005).
\bibitem{Bielewicz05}P. Bielewicz, H. K. Eriksen, A. J. Banday, K. M. G\'{o}rski, and
P. B. Lilje, Astrophys. J. \textbf{635}, 750 (2005).
\bibitem{Vielva04}P. Vielva, E. Mart\'{\i}nez-Gonz\'{a}lez, R. B. Barreiro, J. L.
Sanz, and L. Cay\'{o}n, Astrophys. J. \textbf{609}, 22 (2004).
\bibitem{Cruz05}M. Cruz, E. Mart\'{\i}nez-Gonz\'{a}lez, P. Vielva, and L. Cay\'{o}n,
Mon. Not. R. Astron. Soc. \textbf{356}, 29 (2005).
\bibitem{Cayon05}L. Cay\'on, J. Jin and A. Treaster,
Mon. Not. R. Astron. Soc. \textbf{362}, 826 (2005).
\bibitem{Cruz06a}M. Cruz, Tucci M., E. Mart\'{\i}nez-Gonz\'{a}lez, P. Vielva,
Mon. Not. R. Astron. Soc. \textbf{369}, 57 (2006).
\bibitem{Cruz06b}M. Cruz, L. Cay\'on, E. Mart\'{\i}nez-Gonz\'{a}lez, P. Vielva, and J. Jin,
Astrophys. J. submitted, preprint astro-ph/0603367 (2006).
\bibitem{EMG06}E. Mart\'{\i}nez-Gonz\'{a}lez, M. Cruz, L. Cay\'on, P. Vielva, and J. Jin,
New Astron. Rev. \textbf{in this volumne}, (2006).
\bibitem{Bernui05}A. Bernui, B. Mota, M. J. Rebouças, and R. Tavakol, preprint astro-ph/0511666
(2005).
\bibitem{Hajian03}A. Hajian and T. Souradeep, Astrophys. J. Lett. \textbf{597}, L5 (2003).
\bibitem{Hajian05a}A. Hajian, T. Souradeep, and N. Cornish, Astrophys. J. Lett. \textbf{618},
L63 (2005).
\bibitem{Hajian05b}A. Hajian and T. Souradeep, preprint astro-ph/0501001 (2005).
\bibitem{Hajian06}A. Hajian and T. Souradeep, preprint astro-ph/0607153 (2006).
\bibitem{Jaffe05}T. R. Jaffe, A. J. Banday, H. K. Eriksen, K. M. G\'orski, and F.
K. Hansen, Astrophys. J. Lett. \textbf{629}, L1 (2005).
\bibitem{Jaffe06a}T. R. Jaffe, A. J. Banday, H. K. Eriksen, K. M. G\'orski, and F.
K. Hansen, Astrophys. J. \textbf{643}, 616 (2006).
\bibitem{Jaffe06b}T. R. Jaffe, S. Hervik, A. J. Banday and K. M. G\'orski, 
Astrophys. J. \textbf{644}, 701 (2006).
\bibitem{Cayon06}L. Cay\'on, A. J. Banday, T. R. Jaffe, H. K. Eriksen, F. K. Hansen,
K. M. G\'orski and J. Jin, Mon. Not. R. Astron. Soc. \textbf{369}, 598 (2006).
\bibitem{McEwen06}J. D. McEwen, M. P. Hobson, A. N. Lasenby and D. J. Mortlock,
Mon. Not. R. Astron. Soc. \textbf{369}, 1858 (2006).
\bibitem{Bridle06}M. Bridles, J. D. McEwen, A. N. Lasenby and M. P. Hobson,
Mon. Not. R. Astron. Soc. submitted, preprint astro-ph/0605325, (2006).
\bibitem{Wiaux06a}Y. Wiaux, P. Vielva, E. Mart\'{\i}nez-Gonz\'{a}lez  and P. Vandergheynst, 
Phys. Rev. Lett., \textbf{96},  151303 (2006).
\bibitem{Vielva06}P. Vielva, Y. Wiaux, E. Mart\'{\i}nez-Gonz\'{a}lez  and P. Vandergheynst, 
to be submitted (2006).
\bibitem{Wiaux06b}Y. Wiaux, L. Jacques, P. Vielva and P. Vandergheynst, 
Astrophys. J. accepted, preprint astro-ph/0508516 (2006).
\bibitem{Wiaux06c}Y. Wiaux, L. Jacques, and P. Vandergheynst, J. Comput. Phys. submitted,
preprint astro-ph/0508514 (2006)
\bibitem{Freeman91}W. T. Freeman and E. H. Adelson, IEEE Trans. Pattern Anal. Machine Intell., 
\textbf{13}, 891, (1991).
\bibitem{Wiaux05}Y. Wiaux, L. Jacques, and P. Vandergheynst, Astrophys. J. \textbf{632},
15 (2005).
\bibitem{Komatsu03}E. Komatsu \emph{et al.}, Astrophys. J. Suppl. \textbf{148}, 119 (2003).
\bibitem{Bennet03b}C. L. Bennet \emph{et al.}, Astrophys. J. Suppl. \textbf{148}, 97
(2003).
\bibitem{Gorski05}K. M. G\'orski, E. Hivon, A. J. Banday, B. D. Wandelt, F. K. Hansen,
M. Reinecke, and M. Bartelman, Astrophys. J. \textbf{622}, 759 (2005).
\bibitem{Seljak96}U. Seljak \& M. Zaldarriaga, Astrophys. J., \textbf{469}, 437 (1996)
\bibitem{Gordon05}C. Gordon, W. Hu, D. Huterer, and T. Crawford, Phys. Rev. D \textbf{72},
103002 (2005).
\bibitem{Freeman06}P. E. Freeman, C. R. Genovese, C. J. Miller, R. C. Nichol, and L.
Wasserman, Astrophys. J. \textbf{638}, 1 (2006).
\bibitem{Bennet03c}C. L. Bennet \emph{et al.}, Astrophys. J.  \textbf{583}, 1
(2003).
\bibitem{Turok90}N. Turok and D. N. Spergel, Phys. Rev. Lett. \textbf{64}, 2736 (1990).
\bibitem{EMG90}E. Mart\'{\i}nez-Gonz\'{a}lez and J. L. Sanz, Mon. Not. R. Astron.
Soc. \textbf{247}, 473 (1990). 
\bibitem{Rakic06}A. Raki\'c, S. R\"{a}s\"{a}nen, and D. J. Schwarz, Mon. Not. R. Astron.
Soc. \textbf{369}, 27 (2006).

\end{thebibliography}
\end{document}